# Conductance quantization in etched Si/SiGe quantum point contacts

G. Scappucci,[1,*] L. Di Gaspare,[1] E. Giovine,[2] A. Notargiacomo,[2] R. Leoni [2] and F. Evangelisti[1,2]

[1]*Dipartimento di Fisica "E. Amaldi", Università Roma TRE, V. Vasca Navale 84, 00146 Roma, Italy*

[2]*Istituto di Fotonica e Nanotecnologie, IFN-CNR, Via Cineto Romano 42, 00156 Roma, Italy*

PACS numbers: 73.23.-b ; 73.23.Ad

**Abstract**

We fabricated strongly confined Schottky-gated quantum point contacts by etching Si/SiGe heterostructures and observed intriguing conductance quantization in units of approximately $1e^2/h$. Non-linear conductance measurements were performed depleting the quantum point contacts at fixed mode-energy separation. We report evidences of the formation of a half $1e^2/h$ plateau, supporting the speculation that adiabatic transmission occurs through 1D modes with complete removal of valley and spin degeneracies.

Since the introduction of compositionally graded buffer layers in the strained silicon modulation-doped quantum well layer structure, continuous improvements in the design and optimization of the heterostructure growth parameters have led to the achievement of high mobility also in Si/SiGe two dimensional electron gases (2DEG).[1] The high quality Si/SiGe 2DEG has come out as a promising system for basic research in the field of 2D electron physics, which was previously mainly restricted to GaAs/AlGaAs heterostructures. Significant studies have been reported as the observation of the 2D metal-insulator transition at zero magnetic field[2-4] or the direct measurement of spin and valley splitting of Landau levels in silicon.[5] With mobilities corresponding to mean free paths in the order of the µm, the quality of the Si/SiGe material is adequate to investigate quantum transport phenomena in lower dimensional structures as 1D systems and quantum dots. However, the large majority of 1D conductance investigations have been performed on systems based on GaAs heterostructures. Few works have dealt with the 1D ballistic transport in silicon or Si/SiGe heterostructures. The major reason that has slowed down the progress in strained silicon quantum devices has been the difficulty in obtaining high confinement of charge carriers and an effective gating action. It has been suggested that this is due to leakage currents and parallel conducting path, likely to be caused by dopant segregation at the surface, dislocations and defects inherent in the Si/SiGe heterostructure.[6] Recently, strained-Si has gained considerable interest also for possible applications in the field of quantum information processing.[7] Challenged by the proposal of quantum computing architectures in SiGe quantum dots,[8] different research groups have been exploring alternative fabrication approaches to overcome technological and material related hurdles. Significant progress has been achieved as witnessed by the number of papers published recently that reported a satisfactory gating action on Si/SiGe quantum devices.[9-13]

We previously demonstrated that significant quantum confinement can be achieved by introducing geometrical bends on etched Si/SiGe nanowires and reported the observation of single electron charging effect above 4 K in Si/SiGe single electron transistors[14] and electron magnetic focusing in Si/SiGe quantum cavities.[15] In this paper we investigate the ballistic 1D electron transport in highly confined Si/SiGe heterostructure quantum point contacts (QPC). Since the discovery of conductance quantization in GaAs/AlGaAs systems,[16] QPCs were mostly investigated for fundamental studies. Recently, QPCs are attracting more and more interest also for their functional use as charge sensors capacitively coupled to quantum dots (QD). Notably, a QPC was successfully used as the electrical read-out channel of an individual electron spin in a QD[17] or as the local electrometer in a recent experiment that demonstrated coherent control of coupled electron spins in double QDs.[18]

The QPCs considered in this paper were defined in Si/SiGe heterostructures by etching away the side material and were effectively controlled by a Schottky gate. We report here and discuss the presence at zero magnetic field of a conductance plateau at $\sim e^2/h$ and evidence for a quantization in unit of $\sim e^2/h$, where $e$ is the electron charge and $h$ the Planck's constant.

It is generally accepted that the conductance of one-dimensional ballistic wires is quantized in units $G_0 = 2e^2/h$ when an adiabatic transmission via spin-degenerate modes is taking place.[16,20] In Si/SiGe systems, due to the presence of valley degeneracy for the electrons, it is expected that the conductance would be quantized in multiple integers of $4e^2/h$. Indeed, conductance quantization in units $4e^2/h$ was reported in split-gate quantum point contacts in Si inversion layers[20] and in a Si/SiGe 2DEG,[21] as well as in etched constrictions in SiGe 2DEG.[22] Quite to the contrary, well defined and wide plateaus at multiples of $2e^2/h$ at zero magnetic field were found in nanoscale vertical silicon structures.[23] Authors in Ref. 23 speculate that the narrow size of the conducting channel could be responsible for the reduction of degeneracies, so leading to the observed $2e^2/h$ conductance quantization in Si.

In GaAs systems, the removal of spin degeneracy and the resulting splitting of the conductance plateaus is usually observed by adding an in-plane magnetic field, which causes the Zeeman splitting of the 1D energy subbands. Surprisingly, a conductance quantization in units of approximately $e^2/h$, which appear to lack spin degeneracy even at zero magnetic field, was reported recently in gated carbon nanotubes.[24] Closely related to these findings could be the additional conductance plateau at 0.5-0.7 $G_0$, usually referred to as "0.7 structure". This is a spin-related phenomenon observed at zero magnetic field in clean 1D GaAs systems, originally evidenced by Thomas *et al.*,[25] that has attracted a great deal of attention recently.[26-31] Its presence is assumed to signal the occurrence of non negligible correlation effects, although it does not seem that a general consensus on its origin has been reached as yet.[32-37]

The QPC devices were fabricated on samples containing a high mobility Si/SiGe 2DEG. The 2DEG's are located 70 nm below the surface of Si/SiGe modulation doped heterostructures, grown by chemical vapour deposition. Details of the layer sequence thickness as well as the structural and morphological properties of the 2DEG's are described elsewhere.[38] For the samples considered in this work, a standard analysis of the low-field magnetoresistance at T=300 mK of mesa-etched Hall bars gives an estimate of the 2DEG carrier density $n_{2D}$=9.8x10$^{11}$ cm$^{-2}$, electronic mobility $\mu$=4.1x10$^4$ cm$^2$/Vs and mean free path of ~ 500 nm.

The QPCs were obtained by carving the 2DEG in a double-bend like geometry by electron-beam lithography (EBL) and reactive ion etching with fluorinated gases. The heterostructures were etched to a depth of 100 nm from the surface. In panels (a), (b) and (c) of Fig. 1 we report, respectively, a schematic of the QPC geometry prior to gate deposition, a side-view schematic of the gated QPC and, finally, a scanning electron micrograph of a complete device. The QPC is formed by the narrow conducting channel (width *w*) which originate at the junction between two sections (labelled S and D in Fig. 1(a)) protruding from the outer mesa structure. The S and D sections, 400-nm-wide and 200-nm-long, act as source and drain leads for the QPC. Since the overall dimensions of the constriction are smaller than the mean free path, the electronic transport through the narrow channel is expected to be ballistic. With this approach, on the same 2DEG sample, nanostructures with constrictions of decreasing geometrical width *w* were obtained by reducing the extent of overlap between the S and D sections. As the constrictions become narrow and their effective width comparable with the Fermi wavelength, that in our 2DEG is estimated to be $\lambda_F \sim 50$ nm, they act as quantum points contacts connecting the source and drain. Due to sidewall depletion caused by the surface states generated by the fabrication process, the constrictions have an effective width much smaller than the lithographic one[39] so that the above condition can be easily met even when the lithographic dimension are larger than $\lambda_F$. In this paper we investigate devices with constrictions that measure a lithographic width *w*~160 nm (as the one shown in Fig. 1(c)). We found this width small enough for the constriction to show a clear QPC behaviour in the electronic transport characteristics.

Recent simulations of etched strained-silicon quantum wires with metal gates predicted a large 1D subband separation and capability of the gates in controlling the wire conductance.[40] Challenged by these promising results we adopted for the etched QPC a gating geometry similar to that considered in Ref. 40. A 5/30-nm-thick titanium/gold gate was patterned by EBL and lift-off in the shape of a 100-nm-wide finger gate crossing the etched double-bend. The gate was carefully aligned to within 20 nm with the central constriction. The metal folds along the etched semiconductor surface actually forming a triple Schottky gate for the conducting channel (see Fig. 1(b)). Etched constrictions have strong lateral confining potentials. Also, the surface states completely screen the electric field imposed by the gate on the lateral walls.[40] As a consequence, the gate varies the carrier concentration without affecting the width of the quantum point contact. Therefore, in our devices we can follow the effect of depleting the 1D channel at fixed mode-energy separation.

The leakage from the Schottky gate to the 2DEG was tested on several devices fabricated on different 2DEG chips. At T=450 mK, as the gate voltage was swept from -2 V to +1 V the measured leakage current was smaller than 0.2 pA. This large available working range enables a full control of the conduction through the QPC down to pinch-off. As suggested in Ref 10, the low-leakage level achieved could be due to the small size of the gates, whose active area is less than 100 nm x 160 nm for the devices considered in this work. The deep etch of 100 nm that defines the structures might also play a significant role in reducing the leakage current.

Electronic transport characterisation of the QPC devices was performed at T=450 mK in a custom designed $^3$He refrigerator[41] using standard ac low frequency lock-in techniques. The source-drain excitation (frequency of 17 Hz) was kept as low as 20 µV root mean square to prevent electron heating. The linear-response conductance (i.e. G=dI/dV$_{SD}$ around V$_{SD}$~0) versus the gate voltage V$_G$ is reported in Fig. 2. This is a typical curve we measure in QPC devices with similar geometry. The curve was corrected for a series resistance R$_S$= 19.4 kΩ, originating from both the 2DEG leads and the source and drain contacts. The curve exhibit plateau-like structures close to multiple integers of 0.5 G$_0$. It is worthy of notice that in no case we would be able to subtract a R$_S$ such as to recover plateaus spaced by 1 G$_0$ or 2 G$_0$. The curve was highly reproducible upon cycling V$_G$ from positive to negative voltages or the temperature from 450 mK to 4.2 K or to room temperature. While sweeping the gate voltage we did not observe any hysteresis nor switching event. This is a significant improvement with respect to previous reports on gated Si/SiGe nanowires.[6,14]

Significant information on the ~0.5 G$_0$ (i.e. ~$e^2/h$) plateau has been obtained from the non-linear transport measurements, i.e. the curves of the differential conductance G as a function of finite dc source-drain bias V$_{SD}$ for different gate voltages V$_G$. In Fig. 3(a) we report a series of G-V$_{SD}$ curves, measured in sequence, progressively decreasing the gate voltage from -0.4 to -0.2 V in steps of 2.5 mV. This gate bias range covers the region where the linear conductance reported in Fig. 2 develops the ~0.5 G$_0$ feature. As a preliminary analysis, we point out that for |V$_{SD}$| >10 mV the conductance value of all the G-V$_{SD}$ curves, irrespectively to the gate voltage bias, start to decrease tending toward zero, a clear indication of current saturation. The likely origin of this saturation will be discussed later on. In the $|V_{SD}|$ <10 mV bias range we observe clear asymmetries in the curves, even around zero $V_{SD}$, that we address in terms of a self-gating effect.[27] We correct our data for this electrostatic effect as in Ref. 27 considering only the symmetric combination $G^*(V_{SD}) = \frac{1}{2}[G(+V_{SD})+G(-V_{SD})]$ of the $G(V_{SD})$ traces. Adjacent point averaging was performed to highlight the trend of the data. We report the corrected $G^*(V_{SD})$ curves in Fig. 3(b).

The curves in Fig. 3(b) show an overall evolution very similar to that found in both GaAs quantum point contacts[42] for the $2e^2/h$ quantization and carbon nanotubes for the $e^2/h$ quantization.[24] This evolution can be accounted for by using the single mode contribution of the Landauer theory for each of the plateau seen in Fig. 2. We see in Fig. 3(b) that for large negative values of $V_G$ the conductance is negligible at small $V_{SD}$, meaning that both electrochemical potentials $\mu_L$ (left contact) and $\mu_R$ (right contact) are below the onset energy $E_0$ of the first 1D band. As the negative gate voltage is decreased to -0.34 V [arrow (1)], the electrochemical potential at $V_{SD}$=0 V (i.e. $\mu=\mu_L=\mu_R$) is aligned to the edge of the first 1D band $E_0$, as confirmed by the clear observation of the change of curvature of the neighbouring $G(V_{SD})$ curves. A further decrease of the negative gate bias brings about a rapid increase of the linear conductance (G at $V_{SD} \sim 0$) toward the value ~0.5 $G_0$, corresponding to $\mu$ entering progressively into the $E_0$ band. At $V_{SD} \sim 4$ mV different traces merge at a value close to 0.25 $G_0$ indicating the formation of the half $e^2/h$ plateau, as expected for $E_0$ lying between $\mu_L$ and $\mu_R$.[43] Around $V_G \sim -0.3125$ V, that corresponds to the 0.5 $G_0$ plateau in the G-$V_G$ curve of Fig 2, several curves bundle at 0.5 $G_0$ [arrow (2)]. An interval of $V_G$ values then follows, where the range of conductance equal to 0.5 $G_0$ progressively spreads to higher values of $V_{SD}$, but no contributions to the conductance come from the next mode. Finally, for further reduction of negative $V_G$, the next mode starts to contribute, first for large values of $V_{SD}$, then to the linear conductance. We see indications of the formation of a ~0.75 $G_0$ plateau at $V_{SD} \sim 4$ mV and $V_G = -0.215$ V [arrow (3)].

Finally, we comment on the drastic decrease of conductance for $V_{SD} > \sim 10$ mV. For sufficiently large source-drain bias the bottom of the electron band of the high-energy contact will become higher than the mode onset and, eventually, the electrochemical potential of the low-energy contact will drop below the bottom of the electron band of the high-energy contact. In these conditions the current saturates at a value independent of bias voltage and the differential conductance drops to zero. Another possible effect causing a current saturation is the electron drift-velocity saturation due to carrier heating at large bias and the onset of non-ballistic transport.[22]

In Fig. 3(c) we report the curves of the conductance G versus $V_G$ as measured, in a successive cool-down, at different $V_{SD}$ dc bias that confirm the evolution we have described. The curves at $V_{SD} = 0$ mV and 8 mV provide a clear evidence of the presence in the linear conductance of 0.5 $G_0$ and 1 $G_0$ steps evolving at large $V_{SD}$ to 0.25 $G_0$ and 0.75 $G_0$ structures, respectively. Arrows are a guide for the eyes. In the curve at $V_{SD} = +24$ mV no significant structures appear due to current saturation.

We estimate the energy spacing $\Delta E_{1,0}$ between the first two 1D subbands by analyzing the non-linear conductance curves at fixed gate voltage with the Zagoskin method.[44] In a quantum point contact, when µ lies between the edges of two successive subbands, the subband energy spacing is $\Delta E = e/2(V_1+V_2)$. Here $V_1$ and $V_2$ are the source-drain voltages at which the first two extrema occur in the derivative $dG/dV_{SD}$, i.e. the position of the inflections of the $G(V_{SD})$ curves at fixed $V_G$. Depending on the position of µ below or above the midway between the edges of successive 1D subbands, $V_1$ is a minimum and $V_2$ is a maximum or vice versa. In Fig. 4 we report two representative $dG/dV_{SD}$ curves obtained by numerical differentiation of the curves at $V_G = -0.3375$ V and $V_G = -0.2925$ V of Fig. 3(b). As depicted schematically in the insets, at these gate voltages the electrochemical potential µ lies below and above, respectively, the midway between the first two 1D subbands. Consistently with the relative position of the chemical potential and the band edges suggested, we found that $V_1$ is a minimum and $V_2$ a maximum for the curve at $V_G = -0.3375$ V. The vice versa occurs for the curve at $V_G = -0.2925$ V.

The subband spacing, calculated according to $\Delta E = e/2(V_1+V_2)$, is $\Delta E_{1,0} \sim 4.4$ meV for both curves. This analysis was repeated for other curves, at different gate bias, in which we could mark unambiguously the position of well-resolved extrema. We found that the subband spacing does not vary significantly with the gate voltage. This confirms that, in our quantum point contact, changes in the gate voltage result in a variation of the carrier concentration without altering significantly its width. It is worth emphasizing that, although the overall behaviour of the linear and non-linear conductance upon changing the gate bias can be explained by the single mode contributions of the Landauer theory, a removal of all degeneracies and a quantization in units $e^2/h$ is required to account for the data. The removal of the valley degeneracy is likely to be the result of the strong confining potential, which might split the odd and even states formed by combining the k and –k states at the two minima.[45] Indeed, unambiguous removal of the valley degeneracy was also present in the conductance curves reported in Ref. 23 on etched vertical wires.

More intriguing is the presence of the 0.5 $G_0$ plateau. Although the features are not as well resolved as in the GaAs case due to the much shorter mean free path of electrons in the SiGe heterostructures, we point out the similarity between the present data and those of the "0.7 structure". The "0.7 structure" was originally related to correlation effects involving the electron spin.[25] Since then a great deal of efforts has been dedicated to the understanding of its microscopic origin. One model attributes the effect to a spontaneous spin polarization in the QPC due to exchange interaction.[33,34] Another model[35] claims the formation of a dynamical local moment in the QPC resulting in a spin splitting due to the local Coulomb interaction energy U. This model would account for the observation of many features of Kondo physics in QPC.[29] Other models suggest electron-phonon coupling[36] or Wigner crystallization[32] as source of the effect. The observation we report of an analogous phenomenon in a completely different system like the Si/SiGe QPC is relevant to the problem, since a possible theoretical model is required to be valid also for the material parameters of the Si 2DEG.

Previous investigations on the conductance of Si/SiGe QPC did not find the half $G_0$ quantization. We speculate that a strong confining potential is required in order to have the degeneracy removal and that the techniques adopted in Ref. 20-22 did not provide it. A strong confining potential is present in Ref. 23 and there the conductance curves do show a structure at 0.5-07 $G_0$, although the authors do not mention it. We are currently investigating the relationship between potential strength and shape and the presence of the half $G_0$ quantization.

This work was partially supported by the FIRB project RBNE01FSWY "Nanoelettronica" and the FISR project "Nanotecnologie per dispositivi di memoria ad altissima densità". G. S. thanks A. R. Hamilton for stimulating discussions.

**Figure 1**

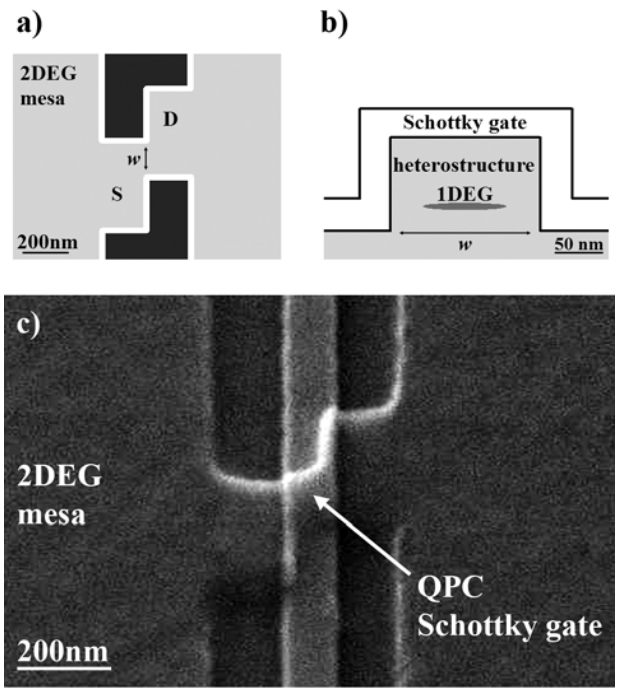

FIG. 1. (a) Top-view schematics of the QPC geometry (prior to gate deposition). The QPC arises in the narrow conducting channel (width *w*) given by the overlap of the S and D sections. (b) Side-view schematics of the etched QPC with the Schottky gate. (c) Scanning electron micrograph of a QPC device at the end of the fabrication process.

**Figure 2**

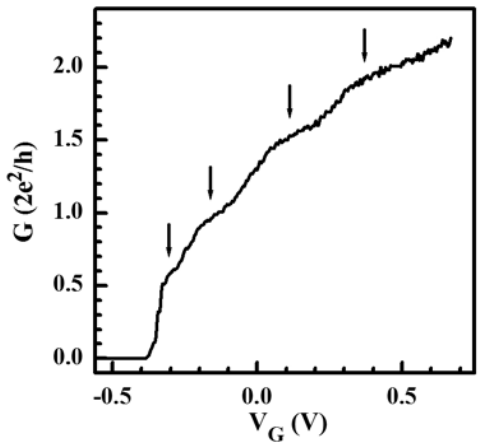

FIG. 2. Differential conductance G versus gate voltage $V_G$ for the device shown in Fig. 1(c) in units of the conductance quantum $G_0=2e^2/h$. This is a two-terminal measurement corrected by a 19.4 kΩ series resistance; measurement temperature was 450 mK.

**Figure 3**

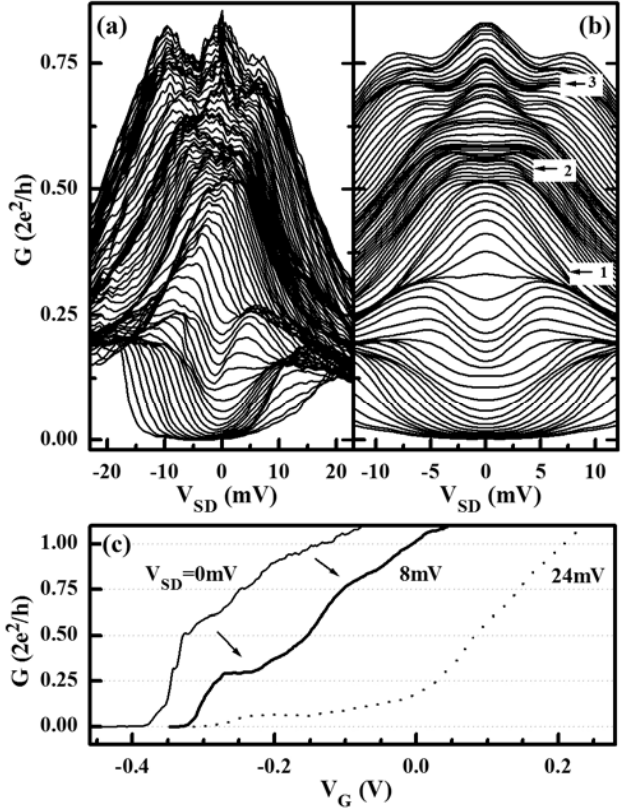

FIG. 3. (a) Plot of the non-linear differential conductance G versus source-drain voltage $V_{SD}$ for different values of gate bias $V_G$ measured at T=450 mK. Both G and $V_{SD}$ are corrected by subtracting a 19.4 kΩ series resistance. Conductance roll-off at $V_{SD}$>~10mV is caused by current saturation. In (b) the symmetrized plot corrected for self-gating effect is reported. Arrows highlight the gate bias values at which significant evolution of the curves is observed, due to the relative alignment between the electrochemical potential µ and the 1D band edges. (c) Differential conductance G versus $V_G$ for three values of $V_{SD}$ bias. The formation of semi-plateau at finite source-drain bias is highlighted by the lowering of the ~1 $G_0$ and ~0.5 $G_0$ structures to ~0.75 $G_0$ and ~0.25 $G_0$ respectively. Traces are offset horizontally. Arrows are a guide for the eyes.

**Figure 4**

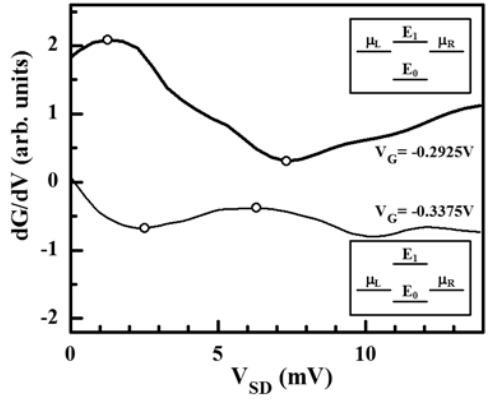

FIG. 4. Conductance derivative dG/dV versus source-drain bias $V_{SD}$ at different gate bias at T=450 mK. The curves are offset vertically. Open circles mark the position $V_1$ and $V_2$ at which the first two extrema occur in each curve. Insets depict the position of the electrochemical potentials $\mu_L$ and $\mu_R$ with respect to the first two 1D subbands.